\documentclass[balance,colorlinks]{asmeconf}

\usepackage{amsmath}
\usepackage{cases}
\usepackage{booktabs}
\usepackage{multirow}
\usepackage{listings}
\newcommand{\tabincell}[2]{\begin{tabular}{@{}#1@{}}#2\end{tabular}} 
\usepackage{graphicx} %
\usepackage{makecell}
\usepackage[strings]{underscore}

\hypersetup{pdfauthor={Qineng Wang, Liming Song, Zhendong Guo, Jun Li, Zhenping Feng}, pdftitle={A Novel Multi-Fidelity Surrogate for Efficient Turbine Design Optimization}}

\begin{document}
\fancyfoot[RO,RE]{Copyright~\textcopyright~2023 by ASME}

\ConfName{Journal of Turbomachinery}
\ConfAcronym{Author manuscript}
\ConfDate{2024, Vol. 146, 041011} %
\ConfCity{} %
\PaperNo{doi:10.1115/1.4064228}

\title{A Novel Multi-fidelity Surrogate for Efficient Turbine Design Optimization} %

\SetAuthors{%
	Qineng Wang\affil{1},
    Liming Song\affil{1},
	Zhendong Guo\affil{1}\CorrespondingAuthor{guozhendong@xjtu.edu.cn}, 
	Jun Li\affil{1}, 
    Zhenping Feng\affil{1}  
	}

\SetAffiliation{1}{Institute of Turbomachinery, Xi'an Jiaotong University, Xi'an, China }

\maketitle
\begingroup\renewcommand{\thefootnote}{}
\footnotetext{Published article: Qineng Wang, Liming Song, Zhendong Guo, Jun Li, Zhenping Feng, ``A Novel Multi-Fidelity Surrogate for Efficient Turbine Design Optimization,'' Journal of Turbomachinery 146(4), 041011 (2024). DOI: \url{https://doi.org/10.1115/1.4064228}. ASME is the original publisher. Copyright \textcopyright\ 2023 ASME.}
\endgroup


\begin{abstract}
To solve the turbine design optimization problems efficiently, surrogate-based optimization (SBO) algorithms are frequently used. To further reduce the cost of turbine design, the multi-fidelity surrogate (MFS) based optimization is proposed by the researchers, who resort to augmenting the small number of expensive high-fidelity (HF) samples by a large portion of low-fidelity (LF) but cheap samples in surrogate modeling and optimization process. Nonetheless, according to our observations, the MFS based optimization sometimes can only have better convergence rate at the early stage of optimization process, but yielding worse final solution than the single-fidelity surrogate (SFS) based optimization that uses high-fidelity samples alone. The reason behind can be explained as follows. With the increase of HF samples in the optimization process, the LF samples can cause negative effect and therefore misleading the optimization search. To address the above issue, an ensemble weighted multi-fidelity surrogate (EMFS) is proposed. Specifically, the density-based spatial clustering of applications with noise (DBSCAN) is used to detect the region where the MFS cannot build a more accurate surrogate, and a local SFS is built there. Then, an EMFS is built by combining the MFS and SFS with adaptive weights, which is used to guide the optimization process. The related algorithm is named as multi- and single-fidelity surrogate fused optimization, i.e., MSFO. Through tests on GE-E3 blade optimization and the film cooling layout design of a turbine endwall, the effectiveness of proposed MSFO is well demonstrated. 

\end{abstract}
\begin{nomenclature}
    \entry{CHT}\;\;\;{Conjugate Heat Transfer}
    \entry{CO}\;\;\;{Core Objective in clustering algorithm}
    \entry{CKO}\;\;\;{Co-Kriging based Optimization}
    \entry{DR}\;\;\;{Density Reachable}
    \entry{DDR}\;\;\;{Directly Density Reachable}
    \entry{EI}\;\;\;{Expected Improvement}
     \entry{EGO}\;\;\;{Efficient Global Optimization with kriging}
    \entry{EMFS}\;{Ensemble weighted Multi-Fidelity Surrogate}
    \entry{LHS}\;\;\;{Latin Hypercube Sampling}
    \entry{MFS}\;\;\;{Multi-Fidelity Surrogate}
    \entry{MSFO}{Multi- and single-fidelity surrogate fused Optimization}
    \entry{SBO}\;\;\;{Surrogate-Based Optimization}
    \entry{SFS}\;\;\;{Single-Fidelity Surrogate}
    \entry{HF}\;\;\;{High-Fidelity}
    \entry{LF}\;\;\;{Low-Fidelity}
    \entry{G}\;\;\;{Global}
    \entry{L}\;\;\;{Local}
    \entry{$\bf{x}$}\;\;\;{Design variable vector}
    \entry{$Y$}\;\;\;{Function prediction that treated as a random variable}
    \entry{$D$}\;\;\;{Dimension of a design optimization problem} 
    \entry{$d(\bf{x})$}\;\;\;{Discrepancy function}
    \entry{$\hat y$}\;\;\;{Surrogate prediction}
    \entry{$s^2$}\;\;\;{Surrogate prediction uncertainty}
    \entry{$\rho$}\;\;\;{Scale factor between the high- and low-fidelity model}
    \entry{$C_{ax}$}\;\;\;{Axial chord length}
    \entry{$p$}\;\;\;{Pressure, Blade pitch}
    \entry{$T$}\;\;\;{Temperature} 
    \entry{$U$}\;\;\;{Velocity} 
    \entry{$\phi$}\;\;\;{Overall cooling effectiveness}
    \entry{$\theta$}\;\;\;{boundary layer momentum thickness}
    \end{nomenclature}

\section{Introduction}
The CFD-based simulations have been more and more widely used for aero-turbine design in the past 20 years~\cite{songResearchMetamodelBasedGlobal2016,ruanVariablefidelityProbabilityImprovement2020,adjeiMultidisciplinaryDesignOptimization2021a,johnsonGeneticAlgorithmOptimization2014}.
On one hand, with the ever-increasing inlet temperature and blade loading of aero-turbines, more advanced CFD techniques are demanded to better capture the sophisticated flow structures and thus evaluate the aero-thermal performance of turbine components with better accuracy~\cite{songOptimizationKnowledgeDiscovery2018,persicoHighFidelityShapeOptimization2019}. 
On the other hand, the cost of CFD simulation per run also increases dramatically with the increase of simulation accuracy, making it rather a challenging task to finish the high-fidelity design optimization of a turbine component within the allowed time and budget~\cite{jolyMachineLearningEnabled2019,lopezGlobalOptimizationTransonic2022,babaeeOptimizationForcingParameters2014}. 

To address the above challenge due to the huge computational cost of CFD simulations, the surrogate-based optimization (SBO) algorithms are frequently used~\cite{forresterRecentAdvancesSurrogatebased2009}, which replaces a large portion of CFD simulations by the surrogate-based cheap approximation in the optimization cycle ~\cite{liuSurveyAdaptiveSampling2017}.
The general process of SBO is as follows~\cite{liuSequentialSamplingGeneration2021a,jonesEfficientGlobalOptimization,baertAerodynamicOptimizationLowPressure2020}.
First, a set of initial training samples are collected by using a space-filling technique such as Latin hypercube sampling (LHS).
Second, a surrogate is built with the training samples.
Third, guided by the trained surrogate, an infill criterion such as expected improvement (EI) is used to search the most promising solution candidate to query, and the new queried sample is added to the training sample set.
The algorithm repeats the second and third steps until the termination condition is met.
\par
Note that the surrogate accuracy can greatly influence the effectiveness of SBO algorithms, but for the aero-thermal design optimization of cooling structures and etc., it can be difficult to collect enough high-fidelity (HF) samples to build an accurate surrogate, particularly at the beginning of the optimization process.
Thereby, the concept of multi-fidelity surrogate (MFS) is proposed by the researchers, who resort to augmenting the small number of HF samples by a large number of low-fidelity (LF) but cheap samples in surrogate modeling and the related optimization process~\cite{parkRemarksMultifidelitySurrogates2017,shiMultiFidelityModelingAdaptive2020}.
Several MFS techniques such as co-kriging, LS-MFS and their variants, etc. have been proposed~\cite{forresterMultifidelityOptimizationSurrogate2007,makkarMachineLearningFramework2022,hanHierarchicalKrigingModel2012}.
Bu et al. used the co-kriging MFS for the cooling layout optimization of a turbine endwall~\cite{buImprovingFilmCooling2022}. 
The conjugate heat transfer (CHT) analysis with the fine mesh is used as the HF model, and the CHT analysis with coarser mesh is used as the LF model. 
While achieving similar final solutions, they found that the computational cost of MFS based optimization is reduced by nearly 40\% when compared to the SFS based optimization with HF samples alone.
Additionally, Kim et al.~\cite{kimHightoLowInitialSample2018} and Zhang et al.~\cite{zhangMultifidelityModelBased2019} used MFS based optimization for the film cooling hole designs, which also demonstrate the advantage of MFS based optimization over the SFS-based optimization. 
\par
However, Lin et al.~\cite{linSequentialSamplingApproach2022a} and Guo et al.~\cite{guoGenerativeMultiformBayesian2022,wangTransferOptimizationAccelerating2020a} observed that the MFS based optimization cannot be guaranteed to achieve better final solutions than that of the SFS-based optimization, though the MFS based optimization can usually have a better convergence rate at the beginning of the optimization process. 
By taking the aerodynamic analysis of a low-pressure turbine blade as an example, Guo et al.~\cite{guoAnalysisDatasetSelection2018} compared the accuracy of MFS and SFS with varied number of HF samples. 
They found that, the MFS does not necessarily to be more accurate than the SFS. Instead, the involvement of LF samples can sometimes cause negative effect, making the accuracy of MFS be worse than that of SFS.
In the meantime, with the increase of HF samples, the SFS is likely to become more accurate than that of MFS. 
This explains why the final solutions of MFS based optimization are sometimes observed to be even worse than those of SFS based optimization. 
\par
Upon the awareness of the above failure of MFS, a set of attempts were made to make full use of the LF samples to improve the overall accuracy of MFS. For instance, through a deep look into the most widely used MFS, i.e., co-kriging, that formulated in the form of ${Y_\text{HF}}({\bf{x}}) = \rho {Y_\text{LF}}({\bf{x}}) + d({\bf{x}})$, Park et al.~\cite{parkRemarksMultifidelitySurrogates2017} argued that the selection of the scale factor $\rho$ between HF and LF models can greatly influence the approximation accuracy of MFS.
More specifically, as shown in their one-dimensional example, when $\rho$ is properly set, the use of LF samples can help to transform the approximating task of a very bumpy curve to the fitting of a simple function such as the linear curve~\cite{parkLowfidelityScaleFactor2018,shuNovelApproachSelecting2019}. 
Then, though the original HF function cannot be well fitted by using SFS with HF samples alone, it can be perfectly approximated using co-kriging.
Along this cue, Zhou et al.~\cite{zhouGeneralizedHierarchicalCoKriging2020} and Bu et al.~\cite{buSelectingScaleFactor2022} proposed different strategies to tune $\rho$ in the co-kriging MFS modeling process, and the effectiveness of their strategies are demonstrated through maximizing the displacement of a micro-aerial vehicle and aero-thermal optimization of a turbine endwall cooling layout, respectively.
\par
Different from the above studies which focus on improving the MFS accuracy, we proposes to make use of both co-kriging MFS and kriging SFS as a weighted ensemble to improve the approximation accuracy of the objective function, and thus realizing the original intention of MFS based optimization, i.e., achieving better final solutions with even fewer cost.
The motivation behind is as follows. 
First, for turbine design problems of which input-output relation cannot be expressed explicitly and analytically, the similarity between the related high- and low-fidelity performance evaluation models is unknown. Thereby, it can be difficult to judge whether or not a proper $\rho$ for ${Y_\text{HF}}({\bf{x}}) = \rho {Y_\text{LF}}({\bf{x}}) + d({\bf{x}})$ can be found to simplify the fitting task of original HF function. 
Second and more importantly, though the LF samples can more or less capture the global trend of the HF objective function, the curve trend of HF and LF models in local regions, particularly that in small neighborhoods of the true optimal solution, may be quite different.
Then, as the iteration goes on especially when a relatively large number of HF samples gathered, the involvement of LF samples can cause negative effect, making the accuracy of MFS be even worse than the SFS that built with HF samples in these promising areas, and thus misleading the optimization search.
Hence, rather than focusing on tuning $\rho$ in ${Y_\text{HF}}({\bf{x}}) = \rho {Y_\text{LF}}({\bf{x}}) + d({\bf{x}})$, we propose to build a local kriging SFS with HF samples in the region where HF samples are relatively densely distributed to guide the optimization process. 
Thereby, the misleading effect of MFS can be greatly alleviated, and the exploitation of better solutions in promising local regions can be also enhanced to accelerate the algorithm optimization progress. 
\par
More specifically, the technique of density-based spatial clustering of applications with noise (DBSCAN) is used to detect the region where HF samples are relatively densely distributed.
Considering the fact that the boundary as whether MFS or SFS can provide more accurate prediction cannot be explicitly defined, an ensemble weighted MFS (EMFS) is proposed, which fuses the predictions of co-kriging MFS and kriging SFS by a sigmoid density function. 
Upon that, a multi- and single-fidelity surrogate fused optimization procedure is proposed, labeled as MSFO, which is used for the aerodynamic optimization of GE-E3 blade and film cooling layout design of a turbine endwall.
\par
With the above, the main contributions of this work can be summarized as follows:

(1) To address the negative effect of LF samples in MFS based optimization for turbine design, a novel ensemble weighted multi-fidelity surrogate (EMFS) is proposed.

(2) Based on EMFS, a multi- and single-fidelity surrogate fused optimization (MSFO) algorithm is proposed for turbine design optimization. 

(3) Through the turbine blade aerodynamic design and endwall film cooling layout design and compared against the classic SFS- and MFS-based optimization algorithms, the proposed MSFO is shown to achieve better solutions with faster convergence rate, and is insensitive to the negative effect of LF samples.
\par
The remainder of the paper is organized as follows. 
A brief introduction of the techniques that used for building the EMFS surrogate is given in Section 2. 
Then, the details of proposed EMFS and the related MSFO algorithm are illustrated in Section 3. 
After that, the optimizations of turbine blade and endwall film cooling layout design by using MSFO are presented in Section 4.
And finally, some conclusions are drawn in Section 5. 

\section{Preliminaries}

In this section, the basics of the techniques such as kriging, co-kriging and DBSCAN that used to build ensemble weighted multi-fidelity surrogate (EMFS) are introduced briefly. 

\subsection{Kriging surrogate}

The kriging surrogate is a popularly used approximation technique~\cite{alizadehManagingComputationalComplexity2020}, which realizes the prediction $Y(\bf{x})$ at an unknown point $\bf{x}$ as a trend function $f(\bf{x})$ plus a Gaussian process $Z(\bf{x})$ as follows:
\begin{equation}
\begin{small}
{Y_{}}({\bf{x}}) = f({\bf{x}}) + Z({\bf{x}})
\end{small}
\end{equation}
Usually, $f(\bf{x})$ is set as a constant that calculated by the following formulation:
\begin{equation}
\begin{small}
\mu=\left(
\mathbf{1}^{T} \mathbf{R}^{-1} \mathbf{1}
\right)^{-1} 
\mathbf{1}^{T} \mathbf{R}^{-1} {\mathbf{y}}_{}
\end{small}
\end{equation}
where, $\bf{l}$ is a $n$-dimensional vector with all elements of one, and $\bf{y}$ is the vector that contains the function values of $n$ samples, and $R$ is the covariance matrix that measures the spatial correlations between the training samples ${\bf{x}}_{i}$ and ${\bf{x}}_{j}$.
Accordingly, the mean prediction $\hat y(\bf{x})$ and related prediction uncertainty $s^2(\bf{x})$ are calculated as follows:
\begin{equation}
\begin{array}{c}
\hat y_\text{SFS}({\bf{x}}) = \mu  + {\bf{r}}{({\bf{x}})^T}{R^{ - 1}}({\bf{y}} - {\bf{l}}\mu )\\
{s^2_\text{SFS}}({\bf{x}}) = {\sigma ^2}\left( {1 - {\bf{r}}{{({\bf{x}})}^T}{R^{ - 1}}{\bf{r}}({\bf{x}}) + \frac{{{{(1 - {{\bf{l}}^T}{R^{ - 1}}{{\bf{r}}}({\bf{x}}))}^2}}}{{{{\bf{l}}^T}{R^{ - 1}}{\bf{l}}}}} \right)
\end{array}
\end{equation}
where, $r(\bf{x})$ is the correlation vector between $\bf{x}$ and the training samples, $\sigma$ is the process variance of samples. More details of kriging can be found in~\cite{jonesTaxonomyGlobalOptimization}.

\subsection{Co-kriging surrogate}

Different from kriging which only uses single-fidelity samples to build approximation, co-kriging combines high- and low-fidelity samples together to build surrogate models, with the following equation:
\begin{equation}
Y_\text{HF}({\bf{x}})=\rho Y_\text{LF}({\bf{x}})+Z_d({\bf{x}})
\end{equation}
where, $Y_\text{HF}$ and $Y_\text{LF}$ are the high- and low-fidelity model, respectively, at unobserved point $\bf{x}$, and $\rho$ is the scale factor between $Y_\text{HF}$ and $Y_\text{LF}$, and $Z_d$ is the discrepancy function that models the difference between the high- and low-fidelity functions. 

Usually, the tuning process of co-kriging is as follows.
First, a low-fidelity surrogate $\hat y_\text{LF}(\bf{x})$ is built by using kriging shown in Section 2.1.
Then, combining the high- and low-fidelity samples. Then, $Z_d$ is also modeled as a Gaussian process, and the scale factor $\rho$ and the hyperparameters of $Z_d$ are estimated by maximizing the related likelihood function. 
Accordingly, the co-kriging prediction and related prediction uncertainty are presented as follows:
\begin{equation}
\begin{aligned}
{{\hat y}_\text{MFS}}({\bf{x}}) &= {\beta _0} + {c^T}({\bf{x}}){C^{ - 1}}({\bf{y}}^{*} - {\beta _0}F)\\
s_\text{MFS}^2({\bf{x}}) &= {\rho ^2}\sigma _\text{LF}^2 + \sigma _d^2 - {c^T}({\bf{x}}){C^{ - 1}}c({\bf{x}})
\end{aligned}
\end{equation}
where, ${\beta _0}$ is the regression constant, $
{{\bf{y}}^*} = [{{\bf{y}}_\text{HF}},{{\bf{y}}_\text{LF}}]^T$ is the vector consisted of the function values of both high- and low-fidelity samples, and $C$ is the covariance matrix of the sample set that including both high- and low-fidelity samples, and $c(\bf{x})$ is the correlation vector between $\bf{x}$ and training high- and low-fidelity samples, and $\sigma_\text{HF}$ and $\sigma_\text{LF}$ are the process variance of low-fidelity and discrepancy samples.
More details of co-kriging can refer to~\cite{forresterMultifidelityOptimizationSurrogate2007}.

\subsection{DBSCAN algorithm}

The DBSCAN is a density-based algorithm for clustering analysis~\cite{schubertDBSCANRevisitedRevisited2017}, which have two important parameters, i.e., $\varepsilon$ and $\text{minPts}$. 
Specifically, $\varepsilon$ denotes the threshold radius that used to define the neighborhood of an sample of a dataset, and $\text{minPts}$ is the minimum number of sample points that closest to an sample within $\varepsilon$-neighborhood.
In the meantime, there are three important concepts in DBSCAN, i.e., core objective (CO), directly density reachable (DDR) and density reachable (DR). 
When the number of points that closest to a sample point (e.g. Point A in Fig.1) within $\varepsilon$-neighborhood is larger than $\text{minPts}$, Point A is treated as a CO.
Furthermore, when Point A is a CO, the points that within the $\varepsilon$-neighborhood are said to be DDR by the Point A,  and two points are DR when they can be connected by a sequence of CO points. 

In Fig.1, the hollow circles are used to present the $\varepsilon$-neighborhood of a dataset.
By setting $\text{minPts}$ to be 4, Point A is a CO. 
Accordingly, the dense circles that closest to Point A within $\varepsilon$-neighborhood are DDR by Point A, and Point B is DR by C, as they can be connected by a sequence of CO points.
In contrast, Point N is neither DDR/DR by Point A, nor it can be DR by Point B or C. 
Hence, the Points A, B, C and related DR points are grouped to a cluster, while Point N is grouped to another cluster. 

Accordingly, the training scheme of DBSCAN is as follows.
First, determine $\varepsilon$ and $\text{minPts}$ for the dataset.
Second, calculate the distances between samples to determine the CO points and the DDR/DR relations between points.
Third, group the points into clusters according to their DDR/DR relations.
In this paper, by grouping the HF samples into clusters and counting the number of HF samples in each cluster using DBSCAN, the regions with dense HF samples are detected, which is used as an indicator to decide whether or not a SFS will be built with HF samples. 
The related details are described in Section 3.1.

\begin{figure}[hbpt]
    \centering\includegraphics[width=0.75\linewidth]{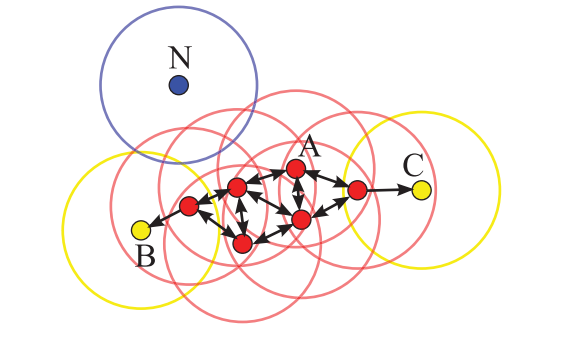}
    \caption{Scheme of the DBSCAN algorithm~\cite{schubertDBSCANRevisitedRevisited2017}\label{fig:1}}
    \end{figure}

\section{Proposed Method}

In this section, the technical details of EMFS and the MSFO algorithm as well are illustrated.

\subsection{Ensemble weighted multi-fidelity surrogate}

The EMFS based prediction and related prediction uncertainty are formulated as follows:
\begin{equation}
\begin{array}{c}
    {{\hat y}_{\text{EMFS}}}({\bf{x}}) = {w_G}({\bf{x}}) \cdot \hat y_\text{MFS}({\bf{x}}) + {w_L}({\bf{x}}) \cdot \hat y_\text{SFS}({\bf{x}})\\
    {{s}^2_{\text{EMFS}}}({\bf{x}}) = {w_G}({\bf{x}}) \cdot s^2_\text{MFS}({\bf{x}}) + {w_L}({\bf{x}}) \cdot s^2_\text{SFS}({\bf{x}})
    \end{array}
\end{equation}
where, kriging that described in Section 2.1 is used to build the local SFS prediction $\hat y_\text{SFS}$, and co-kriging is used to build MFS prediction $\hat y_\text{MFS}$, and $w_G$ and $w_L$ are the weight coefficients of $\hat y_\text{MFS}$ and $\hat y_\text{SFS}$, respectively. 
The subscript $G$ means that $\hat y_\text{MFS}$ is built with all the LF and HF samples over the design space, and $L$ means $\hat y_\text{SFS}$ is built with only HF samples in a local region where HF samples are densely distributed. 
Similar treat for the symbols that used to present prediction uncertainty $s^2_{EMFS}$.
And the training process of EMFS is as follows:

(1) A co-kriging surrogate is built with all the LF and HF samples over the design space. 

(2) The DBSCAN is used to detect the region where HF samples are relatively densely distributed, and accordingly a local kriging with HF samples is built there.

(3) An ensemble weighted surrogate is built, by assigning the weights in Eq.(6) with a sigmoid function.  

In the following paragraphs, the technical details as how to identify the dense distributed area of HF samples and how to assign the weights of adaptively are described in detail.

\subsubsection{Identification of the region with relatively densely distributed HF samples}

On one hand, the surrogate accuracy is highly related to the dense or sparse distribution of samples in a region. 
On the other hand, while the role of LF samples is mainly to capture the global trend of HF function, the local trends of HF and LF models in some promising areas can be different. 
Then, the coupling of LF samples can cause negative effect, making the MFS approximation in these regions be less accurate than a local SFS built with only HF samples, 
In view of the above, the DBSCAN algorithm described in Section 2.3 is used to detect the region where HF samples are relatively densely distributed. 

Specifically, there are two parameters in DBSCAN, i.e., the threshold radius $\varepsilon$ of neighborhood and the minimum number of points $\text{minPts}$ that required in $\varepsilon$-neighborhood to define the CO points and the DDR/DR relations. 
In the meantime, the minimum number of samples that required to build an accurate surrogate is highly related to the problem dimension. 
Moreover, the densely or sparsely distributed samples in a region is a relative concept, and $\varepsilon$ should be related to the number of target data samples in the design space.
Hence, $\varepsilon$ and $\text{minPts}$ are defined as functions shown as follows:
\begin{equation}\label{eq:3.3}
\begin{array}{c}
\varepsilon  = {D^{ - 1/2}}{n_\text{HF}}^{ - 1/D}\\
\text{minPts} = D
\end{array}
\end{equation}
where, $D$ denotes the problem dimension and $n_\text{HF}$ is the number of HF samples.
Then, the process to identify the regions where HF samples are relatively densely distributed is as follows:

(1) Calculate $\varepsilon$ and $\text{minPts}$ with Eq.(7), and use the DBSCAN algorithm to cluster the HF samples.

(2) Based on DBSCAN results, pick out the cluster that contains the maximum number of HF samples, which is the region with relatively densely distributed HF samples, denoted by $S^*$.

(3) Build the kriging SFS with the HF samples in $S^*$, which is the potential to provide more accurate prediction in $S^*$ than that of co-kriging MFS.

\subsubsection{Assigning weights for EMFS}

As pointed out by Liu et al.~\cite{liuSurveyAdaptiveSampling2017}, the density of samples in a local region is linearly correlated with the prediction uncertainty of a Gaussian process, which also measures the prediction accuracy of related surrogate.
In other words, with more densely distributed samples in a region, the prediction uncertainty value of related Gaussian process at an unobserved site $s(\bf{x})$ is smaller, and the related prediction $\hat y(\bf{x})$ is believed to be more accurate, and vice versa.
In the meantime, to accurately measure the density of samples and judge the prediction accuracy of local and global surrogates reasonably, the prediction uncertainty should be calculated by the same Gaussian process with the unified hyper-parameters.
In view of the above, the HF samples are classified into two groups according to the DBSCAN results described in Section 3.1.1.
Wherein, the samples in the DBSCAN cluster that contains the maximum number of HF samples are labeled as $X_L$, and the remaining HF samples will be gathered into another sample group, denoted by $X_G$.
Accordingly, the covariance matrix $\bf{\Sigma}$ and the correlation vector $\bf {r}(x)$ of an unobserved point $\bf{x}$ are calculated as follows:
\begin{equation}
\begin{aligned}
\Sigma =& 
\begin{tiny}
\begin{bmatrix}
{r({\bf{x}}_L^{(1)},{\bf{x}}_L^{(1)})}& \cdots &{r({\bf{x}}_L^{(1)},{\bf{x}}_L^{(m)})}&{r({\bf{x}}_L^{(1)},{\bf{x}}_G^{(1)})}& \cdots &{r({\bf{x}}_L^{(1)},{\bf{x}}_G^{(n)})}\\
 \vdots & \ddots & \vdots & \vdots & \ddots & \vdots \\
{r({\bf{x}}_L^{(m)},{\bf{x}}_L^{(1)})}& \cdots &{r({\bf{x}}_L^{(m)},{\bf{x}}_L^{(m)})}&{r({\bf{x}}_L^{(m)},{\bf{x}}_G^{(1)})}& \cdots &{r({\bf{x}}_L^{(m)},{\bf{x}}_G^{(n)})}\\
{r({\bf{x}}_G^{(1)},{\bf{x}}_L^{(1)})}& \cdots &{r({\bf{x}}_G^{(1)},{\bf{x}}_L^{(m)})}&{r({\bf{x}}_G^{(1)},{\bf{x}}_G^{(1)})}& \cdots &{r({\bf{x}}_G^{(1)},{\bf{x}}_G^{(n)})}\\
 \vdots & \ddots & \vdots & \vdots & \ddots & \vdots \\
{r({\bf{x}}_G^{(n)},{\bf{x}}_L^{(1)})}& \cdots &{r({\bf{x}}_G^{(n)},{\bf{x}}_L^{(m)})}&{r({\bf{x}}_G^{(n)},{\bf{x}}_G^{(1)})}& \cdots &{r({\bf{x}}_G^{(1)},{\bf{x}}_G^{(n)})}
\end{bmatrix}
\end{tiny}\\
=&
\begin{bmatrix}
{{R_{LL}}}&{{R_{LG}}}\\
{{R_{GL}}}&{{R_{GG}}}
\end{bmatrix}
\end{aligned}
\end{equation}

\begin{equation}
{\bf{r}}({\bf{x}}) = 
\begin{bmatrix}
{r({\bf{x}},{\bf{x}}_L^{(1)})}\\
 \vdots \\
{r({\bf{x}},{\bf{x}}_L^{(m)})}\\
{r({\bf{x}},{\bf{x}}_G^{(1)})}\\
 \vdots \\
{r({\bf{x}},{\bf{x}}_G^{(n)})}
\end{bmatrix}
= 
\begin{bmatrix}
{{{\bf{r}}_L}({\bf{x}})}\\
{{{\bf{r}}_G}({\bf{x}})}
\end{bmatrix}
\end{equation}
Then, the prediction uncertainty in the view of the local and global surrogate approximation can be measured by the following equation:
\begin{equation}
\begin{array}{c}
s_L^2({\bf{x}}) = \sigma _L^2\left( {1 - {\bf{r}}_L^T({\bf{x}})R_{LL}^{ - 1}{\bf{r}}_L^{}({\bf{x}}) + \frac{{(1 - {{\bf{l}}^T}R_{LL}^{ - 1}{\bf{r}}_L^{}({\bf{x}}))}}{{{{\bf{l}}^T}R_{LL}^{ - 1}{\bf{l}}}}} \right)
\\
s_G^2({\bf{x}}) = \sigma _G^2\left( {1 - {\bf{r}}_G^T({\bf{x}})R_{GG}^{ - 1}{\bf{r}}_G^{}({\bf{x}}) + \frac{{(1 - {{\bf{l}}^T}R_{GG}^{ - 1}{\bf{r}}_G^{}({\bf{x}}))}}{{{{\bf{l}}^T}R_{GG}^{ - 1}{\bf{l}}}}} \right)
\end{array}
\end{equation}
Accordingly, the weights of local and global surrogates can be formulated as follows:
\begin{equation}
\begin{array}{c}
w({\bf{x}}) = {{s_L^2({\bf{x}})} \mathord{\left/
 {\vphantom {{s_L^2({\bf{x}})} {(s_L^2({\bf{x}}) + s_G^2({\bf{x}}))}}} \right.
 \kern-\nulldelimiterspace} {(s_L^2({\bf{x}}) + s_G^2({\bf{x}}))}}\\
 \\
{w_L}({\bf{x}}) = sigmoid(w({\bf{x}})) = {1 \mathord{\left/
 {\vphantom {1 {(1 + {e^{ - (m \cdot w({\bf{x}}) - b)}})}}} \right.
 \kern-\nulldelimiterspace} {(1 + {e^{ - (m \cdot w({\bf{x}}) - b)}})}}\\
 \\
{w_G}({\bf{x}}) = 1 - {w_L}({\bf{x}})
\end{array}
\end{equation}
where, the $sigmoid$ function is used to enhance the weight of SFS surrogate in the local region where HF samples are relatively densely distributed.
In the meantime, $m$ and $b$ are the parameters of the $sigmoid$ function, which are set as $m=10$ and $b=5$ in this paper.

\subsubsection{Illustration example of EMFS}

To show the effectiveness of the proposed EMFS, a test is conducted for the two-dimensional Ackley function, with the following expression:
\begin{equation}
\begin{aligned}
{f_\text{HF}}({x_1},{x_2}) &=  - 20 \cdot \exp \left( {\sqrt {{{(x_1^2 + x_2^2)} \mathord{\left/
 {\vphantom {{(x_1^2 + x_2^2)} 2}} \right.
 \kern-\nulldelimiterspace} 2}} } \right)\\
 &- \exp \left( {{{(\cos ({{{x_1}} \mathord{\left/
 {\vphantom {{{x_1}} \pi }} \right.
 \kern-\nulldelimiterspace} \pi }) + \cos ({{{x_1}} \mathord{\left/
 {\vphantom {{{x_1}} \pi }} \right.
 \kern-\nulldelimiterspace} \pi }))} \mathord{\left/
 {\vphantom {{(\cos ({{{x_1}} \mathord{\left/
 {\vphantom {{{x_1}} \pi }} \right.
 \kern-\nulldelimiterspace} \pi }) + \cos ({{{x_1}} \mathord{\left/
 {\vphantom {{{x_1}} \pi }} \right.
 \kern-\nulldelimiterspace} \pi }))} 2}} \right.
 \kern-\nulldelimiterspace} 2}} \right) + 39.09\\
{f_\text{LF}}({x_1},{x_2}) &= {f_\text{HF}}({x_1},{x_2})\\ 
&+ 6.8 \cdot (0.585 - 0.00127{x_1} + 0.00113{x_2})\\
{x_i} &\in [ - 2,2]{\rm{}}\;\;\;i = 1,2{\rm{}}
\end{aligned}
\end{equation}

\begin{figure}
\begin{subfigure}[t]{0.25\textwidth} %
\centering{
\includegraphics[width=0.7\linewidth]{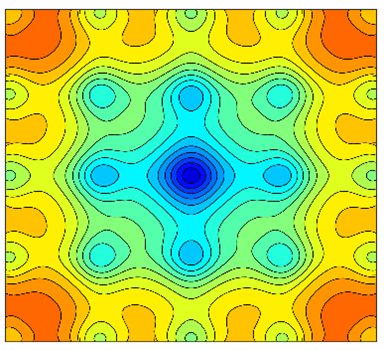}
}%
\subcaption{The contour of the true function}
\end{subfigure}%
\begin{subfigure}[t]{0.25\textwidth}
\centering{%
\includegraphics[width=0.8\linewidth]{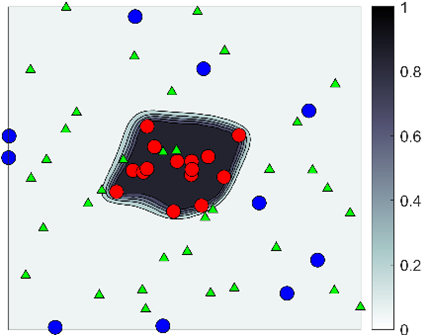}
}%
\subcaption{The contour of SFS weight $w_L$}
\end{subfigure}

\begin{subfigure}[t]{0.5\textwidth} %
\centering{
\includegraphics[width=1\linewidth]{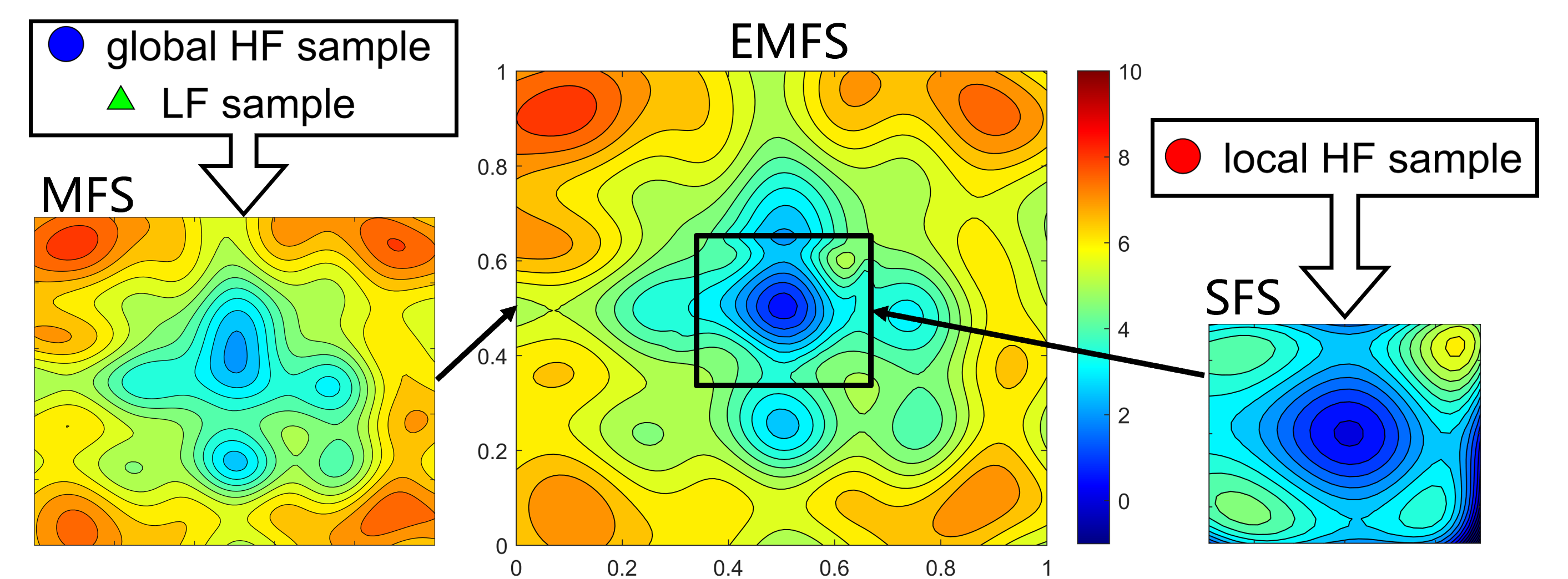}
}%
\subcaption{The contours of MFS, local SFS and EMFS}
\end{subfigure}%
\caption{Illustration of EMFS with two-dimensional Ackley function}\label{fig:3}
\end{figure}

Figure 2(a) shows the contour of true function, which is characterized by bumpy curve with many local optimal solutions. 
To mimic the situation in later stage of the optimization process, the HF and LF samples are drawn from a co-kriging MFS based optimization process that was run several iterations, and 25 HF and 40 LF samples are used for the EMFS modeling.
In Fig.2(b), the LF samples are denoted by triangles, and HF samples are represented by circles.
The LF samples are densely and evenly distributed over the design space.
Guided by the co-kriging MFS in the previous iterations of optimization search, 15 HF samples are densely distributed in a small region with relatively better solutions for a minimization problem (as can be seen by comparing to Fig.2(a)). 
In the meantime, as observed in the leftmost contour of Fig.2(c) that approximated by co-kriging MFS with 25 HF samples and 40 LF samples, the global trend of MFS contour is similar to that of the true function, but the local trend in small neighborhood of the true optimal solution is poorly fitted.
In contrast, as shown in the rightmost contour that built by kriging SFS with 15 HF samples in that region, the neighborhood of the true optimal solution is perfectly approximated. 
It confirms that, in local areas where HF samples are relatively densely distributed, the SFS accuracy can be better than that of MFS.
Then, by combining the MFS and SFS using Eq.(6), the proposed EMFS also accurately captures the local trend that in the vicinity of the true optimal solution, which can then be used to guide the algorithm to achieve the optimal solutions accurately. 
Thereby, the effectiveness of proposed EMFS is well demonstrated.
\par
Also note that, the success of EMFS shown in the above case lies in the following reasons.
First, as the global trend of true HF function can be captured by the co-kriging MFS, the subsequent new queried HF samples in the optimization search are likely to be densely distributed in the promising areas with relatively better objective function values.
Second, the region where HF samples are densely distributed can be accurately detected by the DBSCAN algorithm. 
Last and most importantly, $w_L$ and $w_G$ for EMFS are properly set by our proposed weighting functions shown in Eq.(8) and Eq.(9). 
As can be seen in Fig.2(b), in the region where HF samples densely distributed, $w_L$ is approaching to 1, while it is close to 0 in the remaining areas of the design space that dominated by the MFS.
Thereby, the advantage of the local kriging SFS and global co-kriging MFS can be fully made use of to help the EMFS provide accurate predictions.
As a consequence, the algorithm search guided by EMFS can be expected to take a good balance in between local exploitation and global exploration and thus attaining the optimal solution efficiently. 

\subsection{MSFO algorithm and benchmark test}

Based on EMFS, the details of MSFO algorithm are described in this subsection.
After that, the proposed MSFO is tested on a five-dimensional benchmark function to show the advantage of MSFO over the conventional SFS- and MFS-based optimization procedures.

\subsubsection{MSFO algorithm}
The infill criterion of expected improvement (EI) is used for the optimization search, with the following equation:
\begin{equation}
\begin{aligned}
EI(\bf{x}) &= ({f_{\min }} - \hat y({\bf{x}}))\Phi (u) + s({\bf{x}})\phi ({\bf{x}})\\
u &= {{({f_{\min }} - \hat y({\bf{x}}))} \mathord{\left/
 {\vphantom {{({f_{\min }} - \hat y({\bf{x}}))} {s({\bf{x}})}}} \right.
 \kern-\nulldelimiterspace} {s({\bf{x}})}}
\end{aligned}
\end{equation}
where, $\Phi \left(  \cdot  \right)$ and $\phi \left(  \cdot  \right)$ denote the standard normal distribution function and density function, respectively, and $f_{min}$ is the current best solution, and $\hat y(\bf{x})$ and $s(\bf{x})$ are the function prediction and related prediction uncertainty provided by using co-kriging MFS or EMFS.
\begin{figure}
\centering\includegraphics[width=0.9\linewidth]{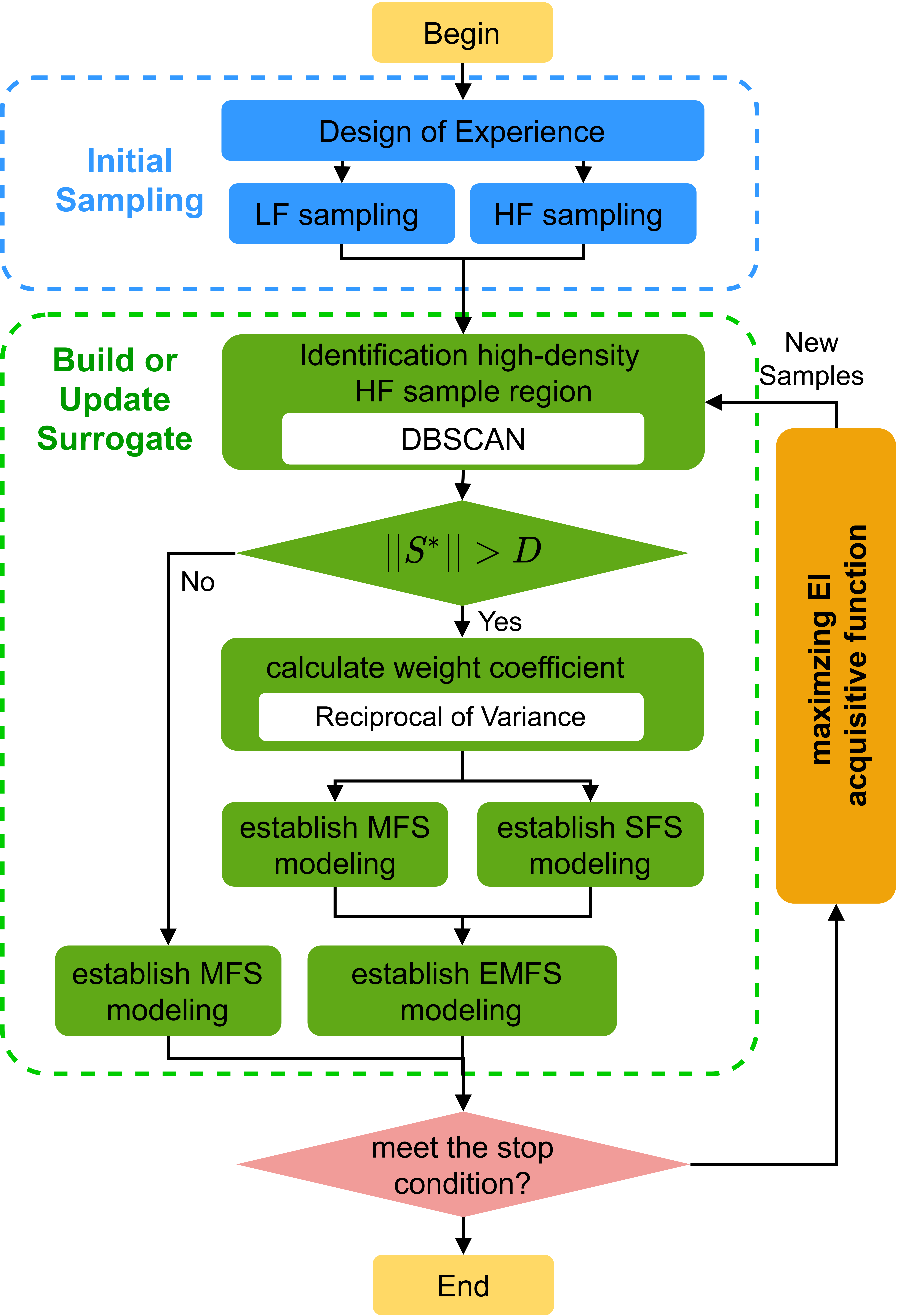}
\caption{The flowchart of MSFO algorithm \label{fig}}
\end{figure}
Accordingly, the flowchart of the MSFO algorithm is shown in Fig.3, which is described as follows:

(1) Design of experiment (DoE) is carried out to generate the initial LF and HF samples.

(2) The DBSCAN is used to detect the region where HF samples are most densely distributed, denoted by $S^*$.

(3) If the number of HF samples in $S^*$ is smaller than the problem dimension $D$, the co-kriging will be built. Otherwise, in addition to building co-kriging MFS, a kriging SFS will be built, and accordingly an EMFS approximation will be built. 

(4) Guided by the co-kriging MFS or EMFS approximation that built in the third step, EI (see Eq.(13)) is used to search the promising solution candidate to query.

(5) Repeat the steps of (2) to (4) until the termination condition is met. 

\begin{figure*}
\begin{subfigure}[t]{0.33\textwidth} %
\centering{
\includegraphics[width=0.9\linewidth]{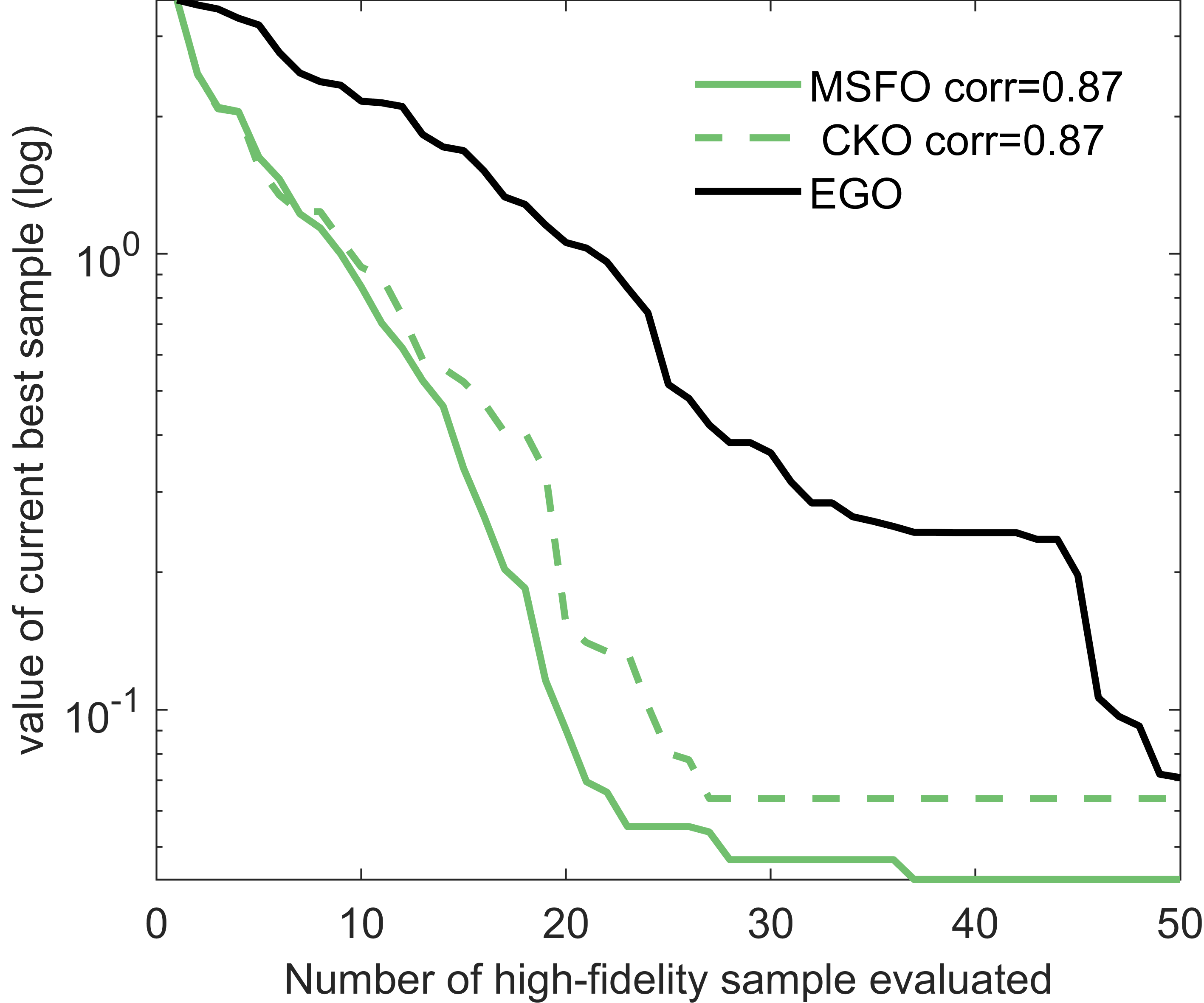}
}%
\subcaption{}
\end{subfigure}%
\begin{subfigure}[t]{0.33\textwidth}
\centering{%
\includegraphics[width=0.9\linewidth]{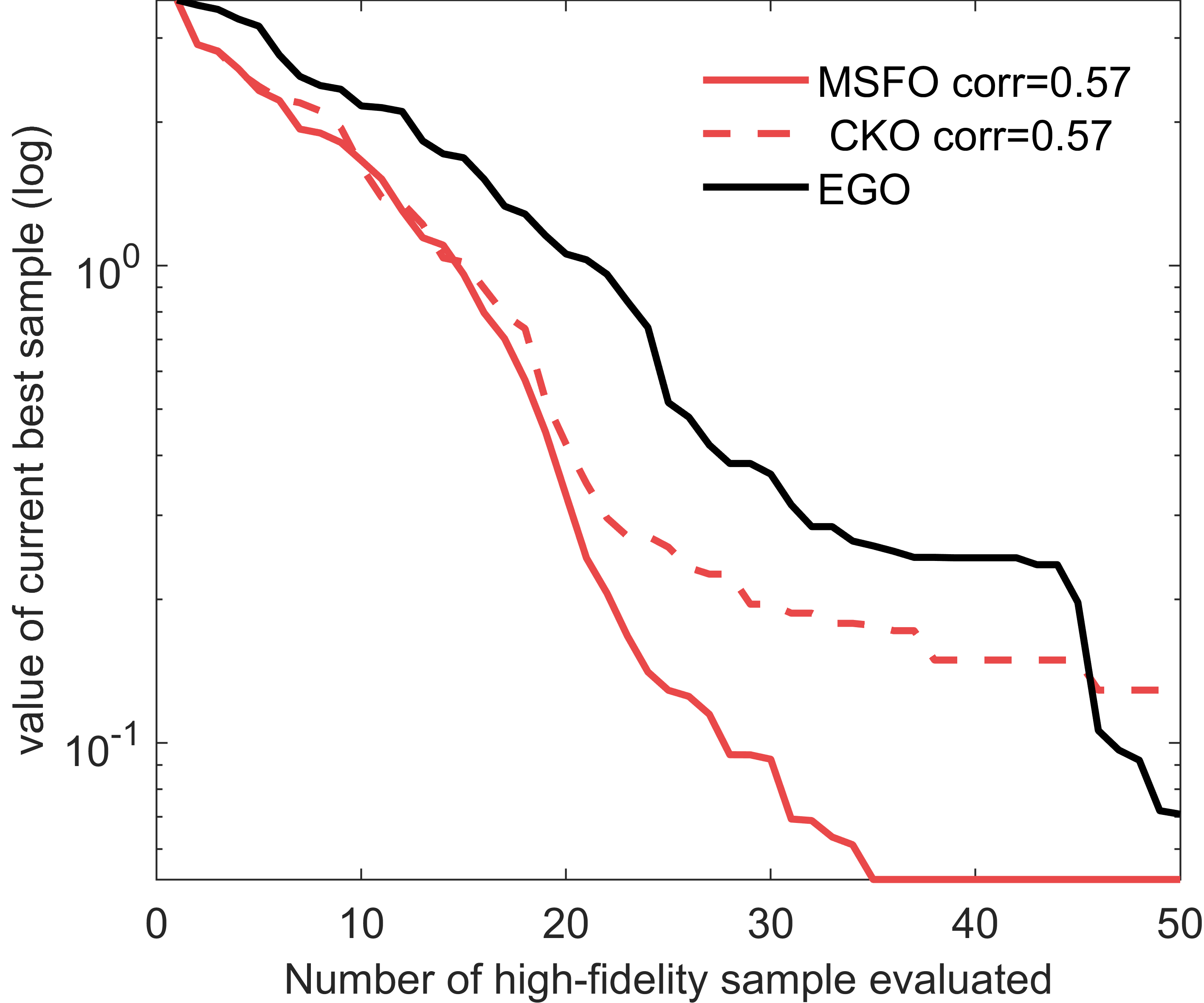}
}%
\subcaption{}
\end{subfigure}
\begin{subfigure}[t]{0.33\textwidth}
\centering{%
\includegraphics[width=0.9\linewidth]{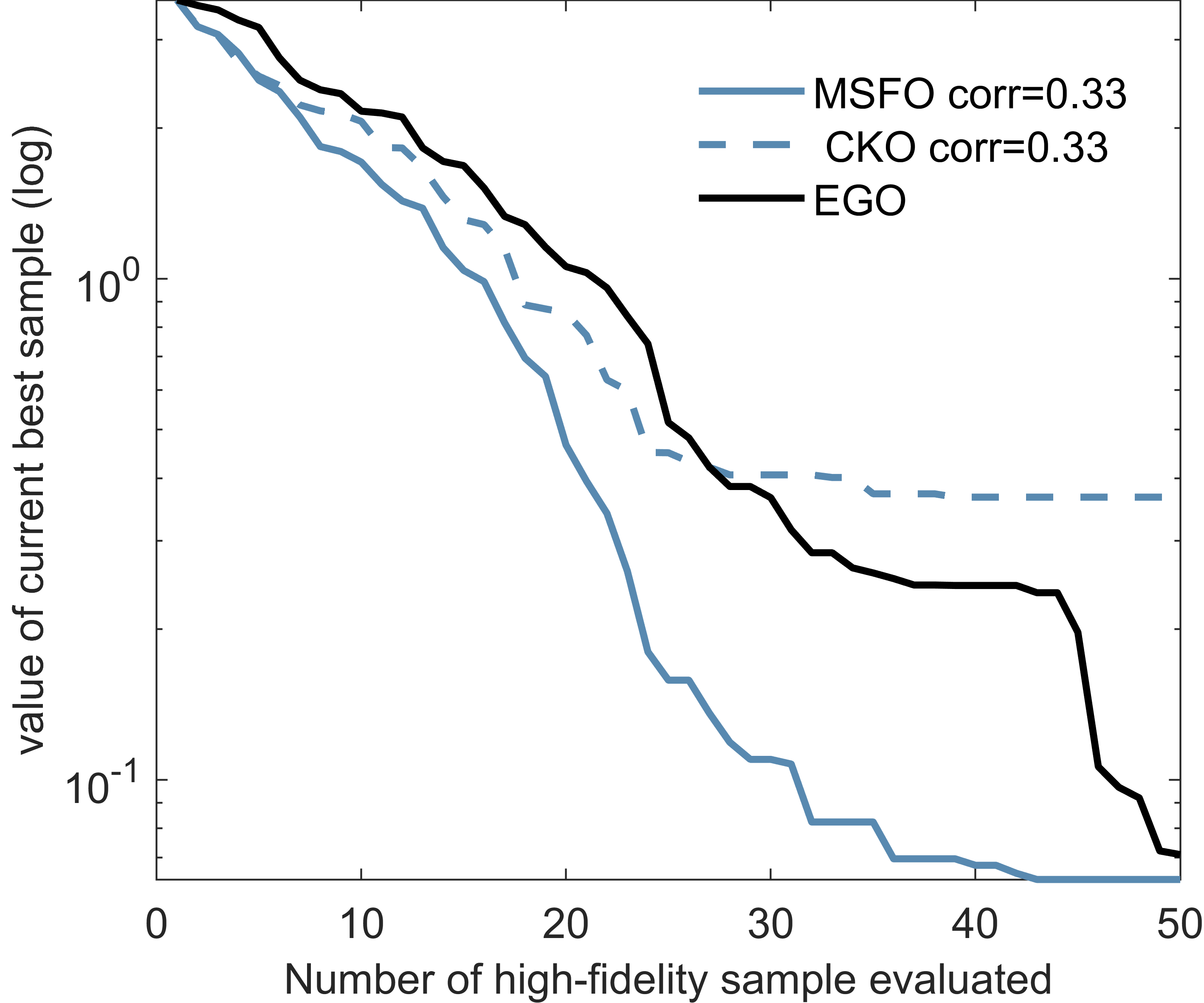}
}%
\subcaption{}
\end{subfigure}

\caption{The averaged convergence curves of five-dimensional Ackley function}\label{fig:4}
\end{figure*}
\subsubsection{Tests on benchmark function}

To show the effectiveness of proposed MSFO algorithm, the five-dimensional Ackley function is used for the test~\cite{zhangVariablefidelityExpectedImprovement2018}, with the following mathematical expression:

\begin{equation}
\begin{small}
\begin{aligned}
{f_\text{HF}}({\bf{x}}) &=  - 20 \cdot \exp \left( {\sqrt {{{\sum\limits_{i = 1}^5 {x_i^2} } \mathord{\left/
 {\vphantom {{\sum\limits_{i = 1}^5 {x_i^2} } 5}} \right.
 \kern-\nulldelimiterspace} 5}} } \right) - \exp \left( {{{\sum\limits_{i = 1}^5 {\cos ({{{x_i}} \mathord{\left/
 {\vphantom {{{x_i}} \pi }} \right.
 \kern-\nulldelimiterspace} \pi })} } \mathord{\left/
 {\vphantom {{\sum\limits_{i = 1}^5 {\cos ({{{x_i}} \mathord{\left/
 {\vphantom {{{x_i}} \pi }} \right.
 \kern-\nulldelimiterspace} \pi })} } n}} \right.
 \kern-\nulldelimiterspace} 5}} \right) \\
&+ 20 + exp(1)\\
{f_\text{LF}}({\bf{x}}) &= {f_\text{HF}}({\bf{x}}) + 6.8 \cdot \alpha \cdot f_{MA5}({\bf{x}})\\
f_{MA5}({\bf{x}}) &= 0.588 - 0.00127{x_1} - 0.00113{x_2} - 0.00663{x_3} \\
&- 0.0129{x_4} - 0.00611{x_5} + 0.00526{x_1}{x_4} + 0.0106{x_1}{x_5}\\
&- 0.000626{x_2}{x_4}- 0.00310{x_2}{x_5} - 0.00724{x_4}{x_5} \\
&- 0.00096x_3^2 - 0.0124x_4^2 - 0.0101x_5^2
\end{aligned}
\end{small}
\end{equation}
Considering the fact that the correlation between $f_\text{LF}(\bf {x})$ and $f_\text{HF}(\bf {x})$ determines whether the LF samples will have positive or negative effect on building accurate MFS and thus influencing the progress of MFS based optimization, 
the parameter $\alpha$ in the expression of $f_\text{LF}$ is set to be 1, 0.6 and 0.1 respectively for the test.
Accordingly, the Pearson's correlation coefficient between $f_\text{LF}(\bf {x})$ and $f_\text{HF}(\bf {x})$ are 0.33, 0.57 and 0.87, respectively, denoted by \begin{small}$corr = 0.33$\end{small}, \begin{small}$corr = 0.57$\end{small} and \begin{small}$corr = 0.87$\end{small}, as shown in the testing plot in Fig.4. 

The algorithms such as EGO and CKO are used as baselines for comparison. 
Specifically, the same infill criterion as EI is used in EGO, CKO and MSFO to search the next promising sample to query.
Differently, EGO uses kriging to build SFS with HF samples alone in the searching process, CKO uses co-kriging to guide the optimization process, while MSFO uses MSFO which combines co-kriging MFS and kriging SFS to guide the optimization search.
At the beginning of the optimization process, the Latin hypercube sampling (LHS) is used to generate the HF and LF samples, and the number of HF and LF samples are set to three times and ten times of the problem dimension, respectively, i.e., $3D$ and $10D$.

Figure 4 shows the mean averaged optimization results over 20 runs.
Specifically, in Fig.4(a), the high- and low-fidelity functions are highly correlated with \begin{small}$corr = 0.87$\end{small}.
In such cases, with the help of LF samples, both CKO and MSFO attain better solutions with faster convergence than EGO. 
In contrast however, when lowering the similarity between the high- and low-fidelity functions as shown in Figs.4(b) and (c), the optimizations with CKO are observed to achieve a faster convergence rate at the beginning of the optimization process, but yielding worse final solutions than those of EGO.
Differently, by making full use of co-kriging MFS and local kriging SFS in a weighted manner as EMFS, the proposed MSFO algorithm can always achieve better final solutions with faster convergence rates. Thereby, the effectiveness of our proposed MSFO algorithm is well demonstrated.

\section{Engineering Optimization}

To show the effectiveness of proposed MSFO for turbine design, aerodynamic optimization of GE-E3 turbine blade and aero-thermal optimization of cooling layout structure of a turbine endwall are carried out in this section. 

Similar to the settings in Section 3.2.2, the LHS is used to generate HF and LF samples at the beginning of the optimization process, and the initial number of HF and LF training samples are set to be $3D$ and $10D$, respectively. 
In the meantime, the EGO and CKO algorithms are used as baselines to show the advantage of proposed MSFO algorithm for turbine design.

\subsection{Aerodynamic Optimization of GE-E3 turbine blade}

The aerodynamic optimization of blade profile of the stator vane of GE-E3 turbine is used as an engineering benchmark for the test.
The mathematical expression of the optimization model is shown as follows:
\begin{equation}
\begin{array}{c}
{{\bf{x}}^*} = \arg \min \xi ({\bf{x}})\\
s.t.{\rm{ }}\;\;0.99m({{\bf{x}}_{ref}}) \le m({\bf{x}}) \le 1.01m({{\bf{x}}_{ref}})
\end{array}
\end{equation}
where, $\xi$ is the energy loss coefficient at the outlet of turbine blade cascade, which is defined as follows:
\begin{equation}
    \varepsilon = 1 - \left[ {1 - {{\left( {\frac{{{P_{{\rm{out }}}}}}{{P_{{\rm{out }}}^*}}} \right)}^{\frac{{\gamma  - 1}}{\gamma }}}} \right]/\left[ {1 - {{\left( {\frac{{{P_{{\rm{out }}}}}}{{P_{{\rm{in}}}^*}}} \right)}^{\frac{{\gamma  - 1}}{\gamma }}}} \right]
\end{equation}
where, $\gamma$ denotes the adiabatic coefficient, and $P_{{\rm{in}}}^*$ and $P_{{\rm{out }}}^*$ are the total pressure at the cascade inlet and outlet, respectively,
and $P_{{\rm{out }}}$ are the static pressure at the cascade outlet.

The modified Pritchard geometric parameterization method~\cite{pritchardElevenParameterAxial1985,agromayorUnifiedGeometryParametrization2021} is used for the parameterization of the blade profile, as shown in Fig.5, and seven design variables are selected for the blade profile optimization, as shown in Table 1. 

\begin{figure}[htbp]
\centering\includegraphics[width=0.65\linewidth]{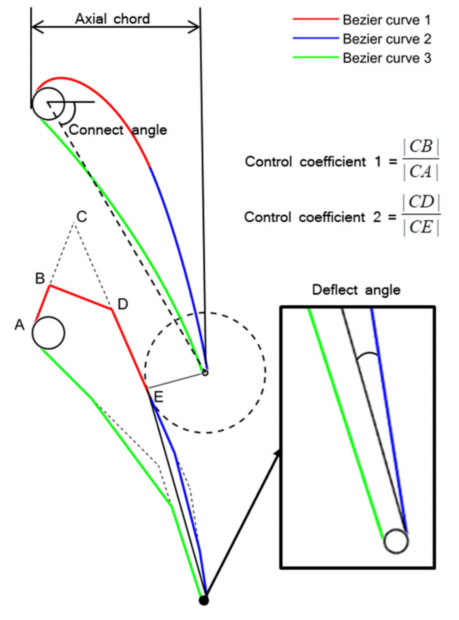}
\caption{Parameterization of GE-E3 blade profile\label{fig:5}}
\end{figure}

\begin{table}[htbp]
    \caption{Variables for blade profile optimization}
    \begin{center}
    \renewcommand\arraystretch{1.2}
    \begin{tabular}{cccc}
    \hline
        \tabincell{c}{No.} &  \tabincell{c}{Geometric\\ definition} & \tabincell{c}{Reference\\value }& \tabincell{c}{Variation\\ range} \\
    \hline
            1 & axial chord[mm] &      33.9  &     (-2.0, 2.0) \\
            2 & center connect angle[$^\circ$] &       59.8  &     (-4.0, 2.0) \\
            3 & inlet wedge angle[$^\circ$] &         69.0 &    (-15.0, 3.0) \\
            4 & outlet deflect angle[$^\circ$] &        4.5  & (-1.5, 4.5) \\
            5 & correlation coefficient[-] &       0.35  & (-0.05, 0.10) \\
            6 & control coefficient 1[-] &        0.40 & (-0.15, 0.15) \\
            7 & control coefficient 2[-] &        0.50  & (-0.15, 0.15) \\
    \hline
    \end{tabular}  
    \end{center}
    \end{table}

The commercial software CFX 15.0 is used for the performance evaluation of the blade profile.
The CFD evaluations with fine and coarse meshes are used as high- and low-fidelity models, respectively, for blade profile optimization.
For benchmark testing purposes, one fine mesh with 23377 nodes and two coarse meshes with 2890 and 1155 nodes, respectively, are used for the CFD evaluation of the blade profile.
Accordingly, the Pearson's correlation coefficient between the fine mesh with 23377 nodes and the coarse mesh with 1155 nodes is 0.79; and that between the fine mesh and the coarse mesh of 2890 nodes is 0.9. 
\begin{figure}[htbp]
\centering\includegraphics[width=0.8\linewidth]{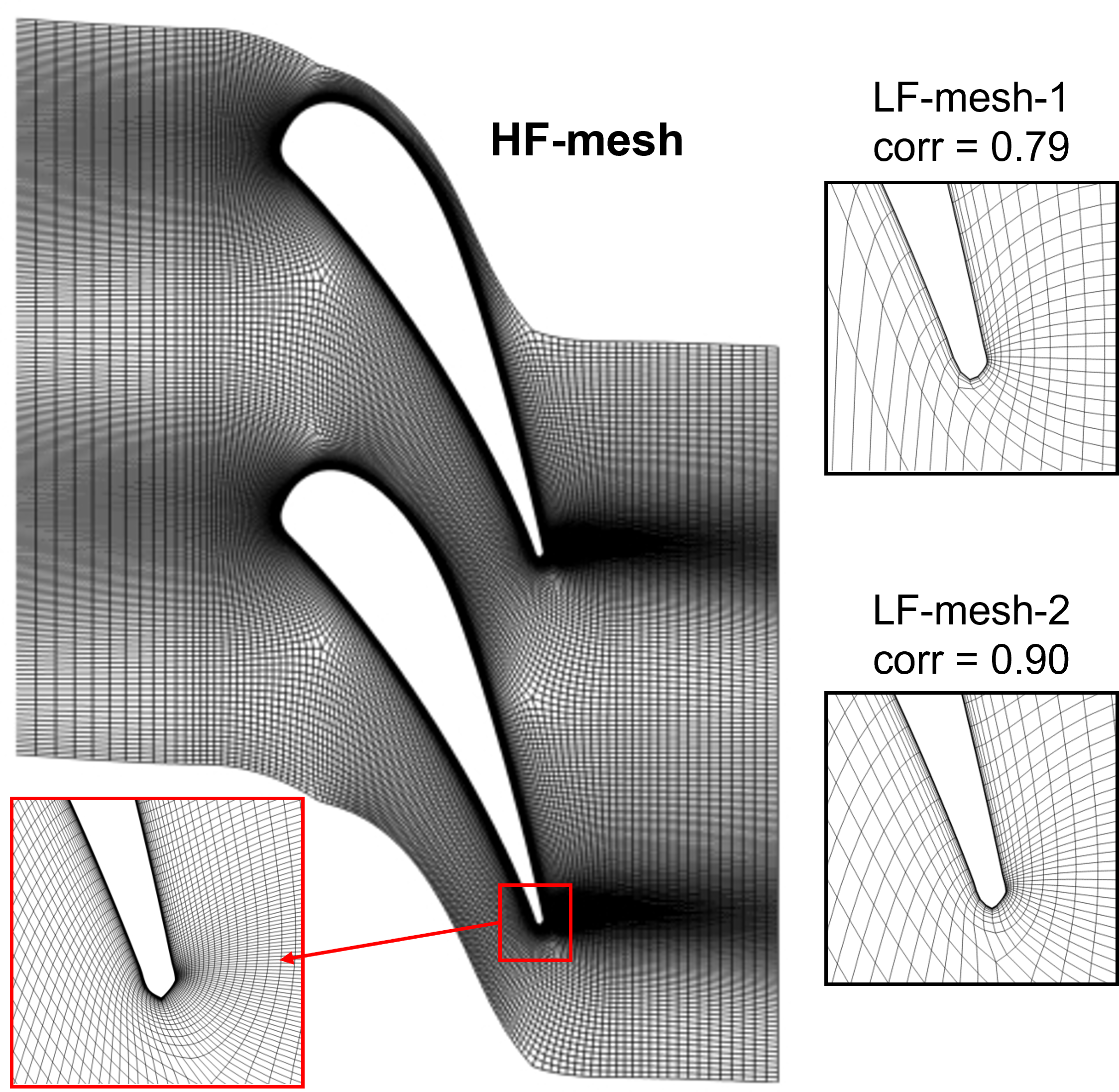}
\caption{Fine and coarse meshes for the high- and low-fidelity performance evaluation of the blade profile.\label{fig:6}}
\end{figure}
The averaged optimization results over 10 runs are shown in Fig.7.
Similar to the benchmark function testing results shown in Fig.4, the CKO is observed to have better convergence rate in the early stage of the optimization process, but attains worse final solution than that of EGO.
\begin{figure}[htbp]
\centering\includegraphics[width=0.8\linewidth]{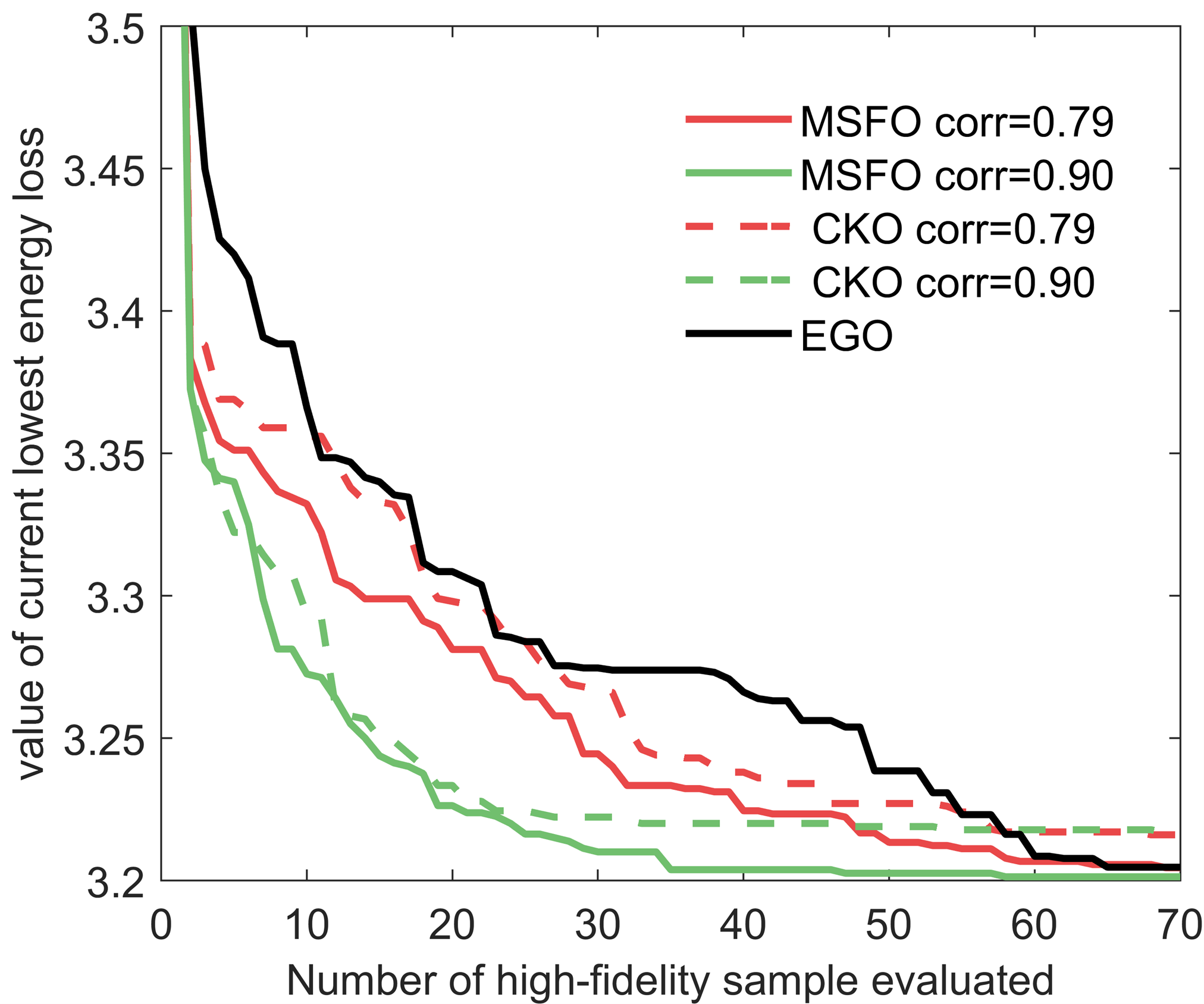}
\caption{The averaged convergence curves of GE-E3 blade profile optimization\label{fig:7}}
\end{figure}
Differently, by building a local SFS to assist the MFS based optimization in an ensemble weighted manner, our proposed MSFO can always achieve similar final solutions to that of EGO with much faster convergence rates. 
Thereby, the effectiveness of the proposed MSFO is demonstrated.
\begin{table}[htbp]
  \centering
  \caption{The detailed optimal result of GE-E3 optimization task}
    \begin{tabular}{ccccc}
        \hline
    \multicolumn{2}{c}{Algorithm}    & \tabincell{c}{Energy\\ loss [\%]} & \tabincell{c}{ Mass flow rate\\ {[$ kg\, s ^{-1}\, m^{-1}$]}}  & \tabincell{c}{ Outlet\\ angle{[$^\circ$]}} \\
  \hline
    \multicolumn{2}{c}{Baseline} & 3.980 & 340.8  & 74.85  \\
    \multirow{2}[0]{*}{MSFO} & \multicolumn{1}{l}{corr=0.79} & 3.204 & 340.3  & 74.92  \\
          & \multicolumn{1}{l}{corr=0.90} & 3.200 &340.2  & 74.93  \\ 
    \multirow{2}[0]{*}{CKO} & \multicolumn{1}{l}{corr=0.79} & 3.223 & 340.5  & 74.91  \\
          & \multicolumn{1}{l}{corr=0.90} & 3.221 & 339.8  & 74.94  \\
    \multicolumn{2}{c}{EGO} & 3.208 & 340.3  & 74.92  \\
    \hline
    \end{tabular}%
  \label{tab:addlabel-1}%
\end{table}%
For aerodynamic validation purpose, Table 2 compares the aerodynamic performance of the medians of the optimal solutions of EGO, CKO and MSFO.
Compared to the original GE-E3 blade profile, the energy loss coefficient of the median optimal solutions of EGO, CKO and MSFO are reduced by , respectively.
In the meantime, the mass flow rates of all optimal solutions meet the mass flow constraints. 
Furthermore, Fig.8 shows the blade loading distributions of GE-E3 blade and the median optimal solutions.
Compared to the loading distribution of original GE-E3 blade, the optimal solutions of EGO, CKO and MSFO display the characteristics of aft-loaded.
As the aft-loaded like blade loading distributions is helpful to reduce the fraction loss of the blade profile, the reason why the optimal solutions achieve better aerodynamic performance can be well understood.

\begin{figure}[htbp]
\centering\includegraphics[width=0.7\linewidth]{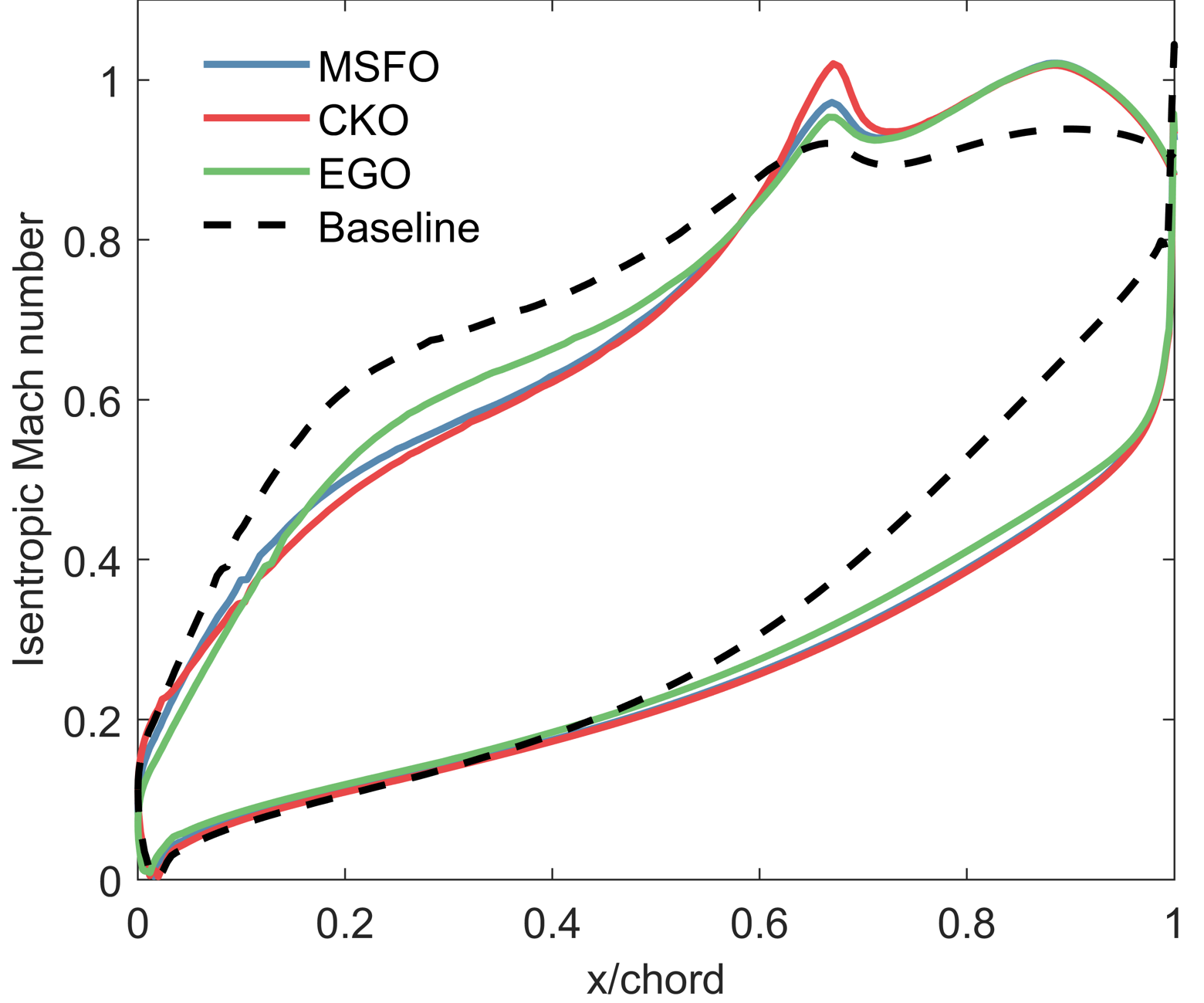}
\caption{Comparison of blade loading distributions of the GE-E3 blade and optimal solutions\label{fig:8}}
\end{figure}

\subsection{Turbine endwall cooling layout design problem}

Upon the aerodynamic benchmark test shown in Section 4.1, our proposed MSFO is further used for the cooling layout optimization with CHT method.
Specifically, the geometry that has been experimentally validated by Mensch and Thole~\cite{menschEffectsNonaxisymmetricEndwall2016} is used in this study. As shown in Fig.9, there are 10 film cooling holes on the endwall with an injection angle of 30$^\circ$, and the compound angles are determined by the local streamline direction, ranging from 20$^\circ$ to 86$^\circ$, and the diameters of the film holes are 4.4 mm.
In the meantime, the inlet of the simulation model for the mainstream is set as 3.5 $C_{ax}$ of the blade leading edge, and the outlet is set to be 1.5 $C_{ax}$ of the blade trailing edge.
More detailed specifications of the geometry and boundary parameters are shown in Table~\ref{tab:addlabel-2}. 
\begin{table}[htbp]
    \centering
    \caption{Specifications of the computational model}
      \begin{tabular}{lclc}
        \hline
      \multicolumn{2}{c}{Geometry parameter} & \multicolumn{2}{c}{Boundary condition} \\
      \hline
      name  & \multicolumn{1}{l}{value} & name  & \multicolumn{1}{l}{value} \\
      $C_{ax}$[mm]  & 218   & $U_{\infty}[m\, s^{-1}]$ & 10.5 \\
      $p/C_{ax}$[-]     & 0.826 & $T_{\infty}$[K]   & 323 \\
      $s/C_{ax}$[-]     & 2.5   & $T_{c}$[K]   & 283 \\
      $\alpha_{in}$[$^\circ$] & 35    &   $\theta[s^{-1}]$    & 0.0046 \\
      $\alpha_{out}$[$^\circ$] & 60 & $m[kg\,s^{-1}]$ & 0.00108 \\
      \hline
      \end{tabular}%
    \label{tab:addlabel-2}%
  \end{table}%

\begin{figure}[htbp]
\centering\includegraphics[width=0.9\linewidth]{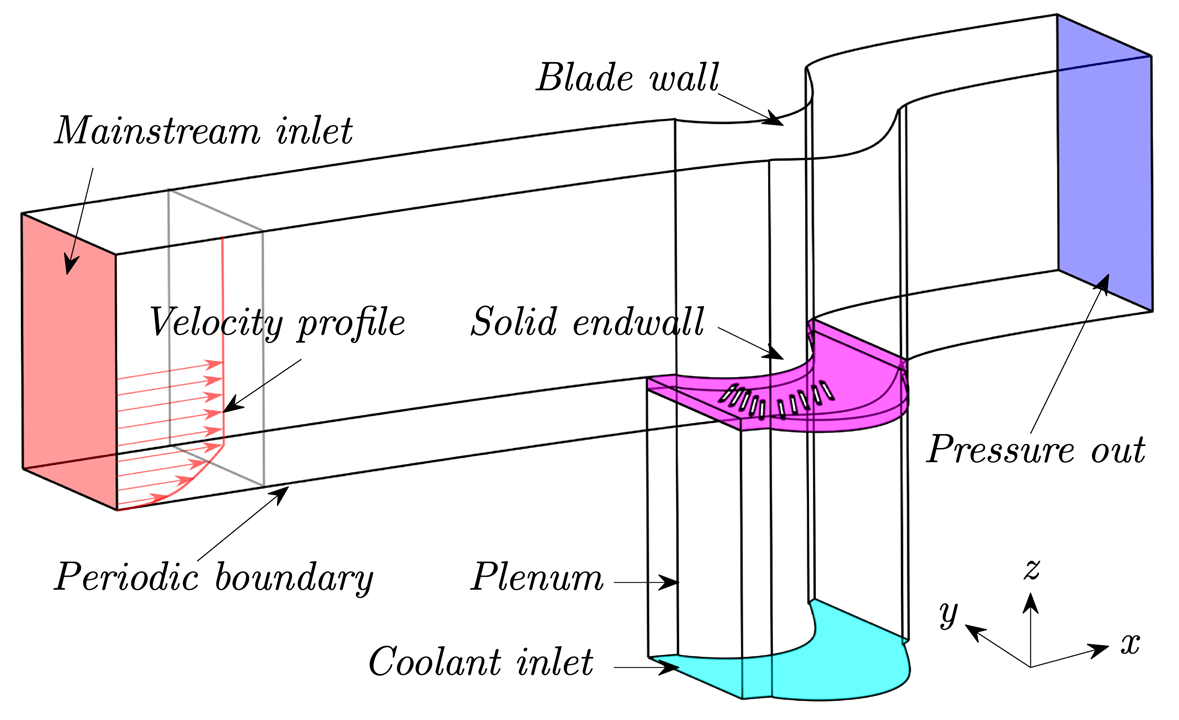}
\caption{Computational model of the turbine endwall cooling layout~\cite{buImprovingFilmCooling2022}}
\end{figure}

The commercial software STAR-CCM+ is adopted, and the $v^2-f$ turbulence model is used for the aero-thermal performance evaluation of endwall cooling layout, the detailed settings of boundary conditions and related numerical validation results can refer to our previous work~\cite{buImprovingFilmCooling2022}.

For the endwall cooling layout design optimization, the fine mesh with 6.35 million nodes that has been experimentally validated in ~\cite{buImprovingFilmCooling2022} is used as the HF evaluation model, and the computational cost per run would take 4 hours when implementing it on a  micro server with two AMD EPYC 7502 32-Core Processor (2.50 GHz) and 256 GB RAM.
In the meantime, to reduce the computational cost, the LF evaluation model is run with a coarser mesh of 0.8 million nodes, and the single run would take 40 minutes.
Accordingly, the Pearson's correlation coefficient between the HF and LF models are 0.87.

The positions of cooling holes are taken as design variables to be optimized. 
As shown in Fig.10, each hole is defined by two parameters, i.e., $u$ and $v$, which are the normalized axial and circumferential coordinates of cooling holes.
For the circumferential row of cooling holes, the distance between the holes in circumferential direction are kept equidistant, thereby the position change of the circumferential holes can be adjusted by varying $v_{c1}$ and $v_{c5}$, i.e., the circumferential parameters of the first and the last circumferential hole.
For the five axial row of cooling holes, the distances between them in the axial direction are kept equidistant.
As the first axial hole is very close to the holes that are arrayed in the circumferential direction, it only varied along the circumferential direction with variable $v_{a1}$.
In the meantime, the last axial hole is moved in both axial and circumferential direction by adjusting variables $u_{a5}$ and $v_{a5}$.
Thereby, five variables are used for the endwall cooling layout optimization, and their range are shown in Table 4.

Also note that, in addition to changing the positions of the cooling holes, the direction of compound angle is also changed, which is adjusted by assuming that the injection direction of the coolant is in the plane that parallel to the tangential direction of the local virtual streamline, as shown by the dashed lines in Fig.11. This way, the interference between film cooling hole pipes can be prevented naturally.

\begin{figure}[htbp]
\begin{subfigure}[t]{0.25\textwidth} %
\centering{
\includegraphics[width=0.8\linewidth]{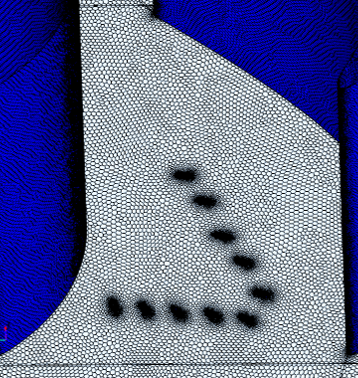}
}%
\subcaption{HF model with fine mesh}
\end{subfigure}%
\begin{subfigure}[t]{0.25\textwidth}
\centering{%
\includegraphics[width=0.8\linewidth]{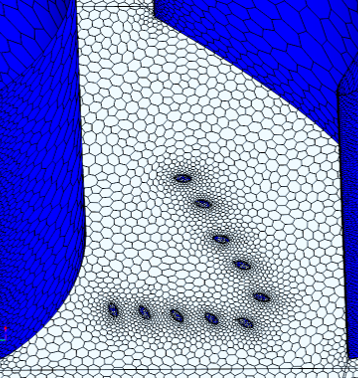}
}%
\subcaption{LF model with coarse mesh}
\end{subfigure}
\caption{Fine and coarse meshes for high- and low-fidelity aero-thermal performance evaluation of endwall cooling layout}\label{fig:10}
\end{figure}

\begin{figure}[htbp]
\centering\includegraphics[width=1.0\linewidth]{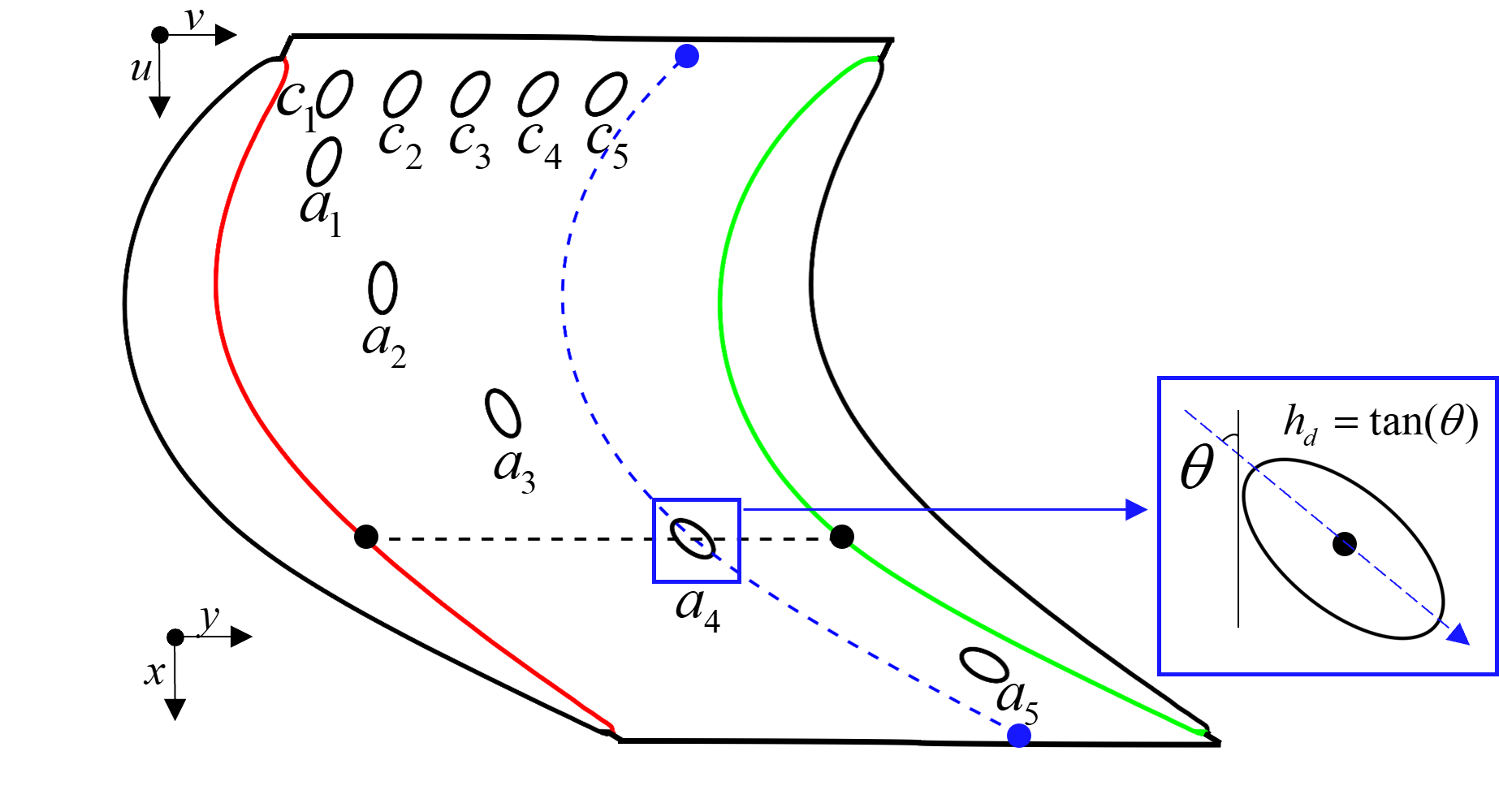}
\caption{The definition of parameters in the optimization of endwall cooling layout design\label{fig:9}}
\end{figure}

\begin{table}[htbp]
    \centering
\caption{Variables for cooling layout optimization}\label{tab:2}
    \centering
    \begin{tabular}{ccc}
    \hline
    No. &Design Variable & Variation Range  \\ \hline
        1 &$v_{c1}$ &  [0.05,0.30]  \\ 
        2 &$v_{c5}$  & [0.70,0.93]  \\ 
        3 &$v_{a1}$  & [0.05,0.93]  \\ 
        4 &$v_{a5}$  & [0.05,0.93]  \\
        5 &$u_{a5}$ &  [0.55,0.90]  \\ \hline
    \end{tabular}
\end{table}

\begin{figure}[htbp]
\centering\includegraphics[width=0.7\linewidth]{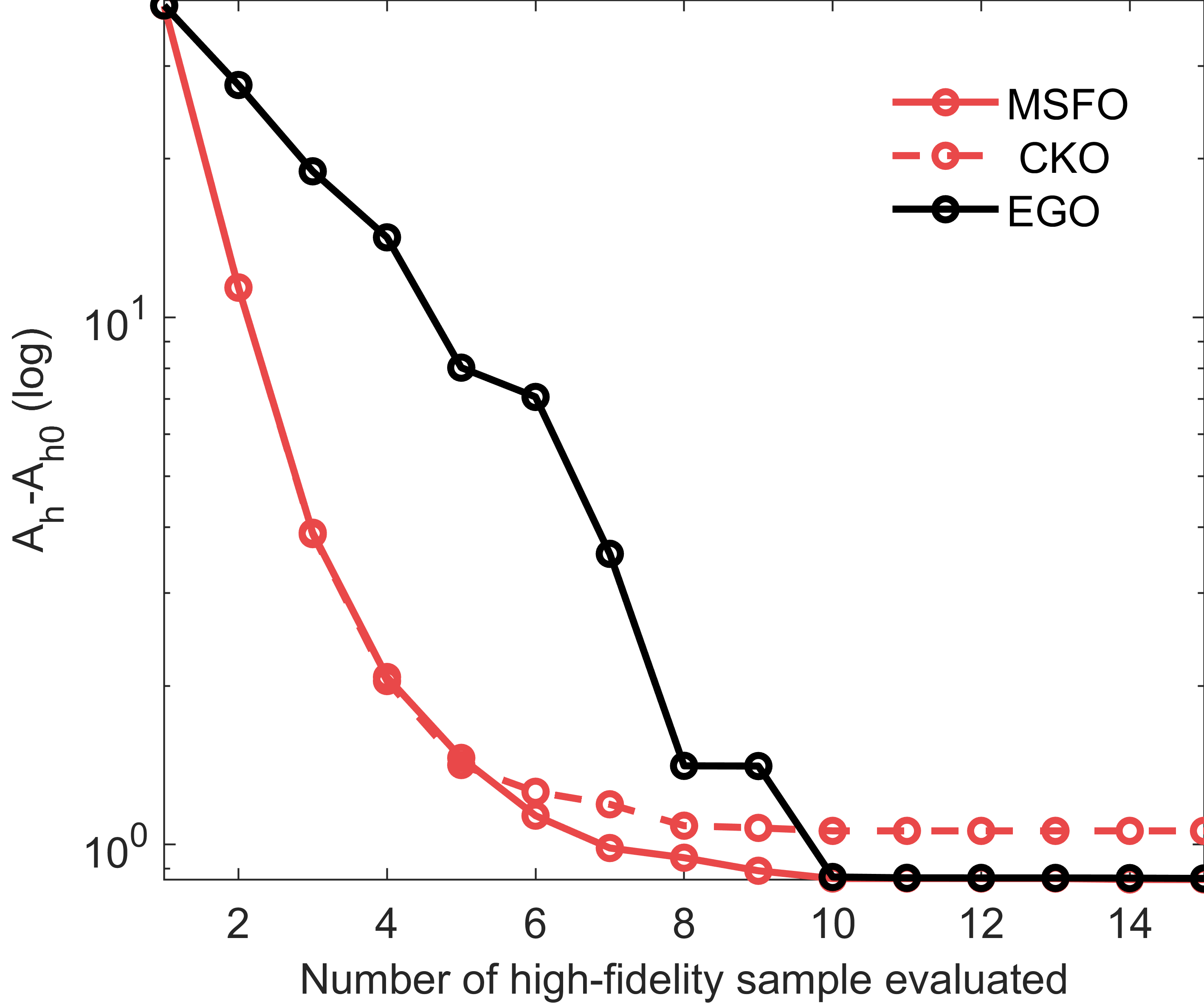}
\caption{The convergence curves of the endwall layout optimization\label{fig:11}}
\end{figure}

By following the settings in~\cite{buImprovingFilmCooling2022, youngDefiningTheEfficiencyCooled2006}, the objective function of cooling layout optimization is set as minimizing the overheating area $A_h$ with the following equation:
\begin{equation}
{{\bf{x}}^*} = \arg \min {A_h}({\bf{x}})
\end{equation}
where, $x$ is the design variable vector that contains the variables shown in Table 4, and $A_h$ is defined as follows:
\begin{equation}
{A_h} = \int_{s:\phi  < \phi_{cr} } {ds} 
\end{equation}
where, $\phi$ is the overall cooling effectiveness, which is defined as $\phi  = {{({T_\infty } - {T_w})} \mathord{\left/
 {\vphantom {{({T_\infty } - {T_w})} {({T_\infty } - {T_c})}}} \right.
 \kern-\nulldelimiterspace} {({T_\infty } - {T_c})}}$, and $T_\infty$ is the temperature of the incoming flow, and $T_w$ and $T_c$ are temperature of the wall and coolant, respectively.
In the meantime, $\phi_{cr}$ is the critical $\phi$ which is set to be 0.15 in this case to detect the overheat areas.

Figure 12 shows the optimization results of the endwall cooling layout optimization, by using MSFO, CKO and EGO, respectively.
Similar to the situations in benchmark function test and turbine blade aerodynamic design, the conventional MFS based optimization algorithm CKO is observed to achieve better convergence rate at the early of the optimization stage, but it attains worse solution than that of EGO, which uses kriging SFS to guide the optimization process.
Differently, by building kriging SFS and co-kriging MFS simultaneously and combine them in a ensemble weighted manner to guide the optimization process, our proposed MSFO achieves similar optimal solutions as EGO algorithm with much faster convergence rate.

\begin{figure}[htbp]
    \centering
    \begin{subfigure}[c]{0.4\textwidth}
        \centering
        \includegraphics[scale=0.6]{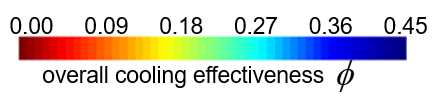}
        \subcaption*{ }
    \end{subfigure}

    \begin{subfigure}[c]{0.4\textwidth}
        \centering
        \includegraphics[scale=0.8]{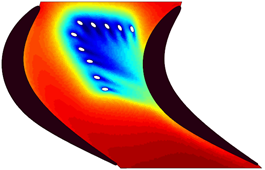}
        \subcaption{Reference design}
    \end{subfigure}

    \begin{subfigure}[c]{0.4\textwidth}
        \centering
        \includegraphics[scale=0.8]{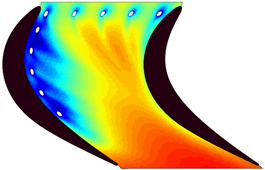}
        \subcaption{Optimal solution of the MSFO}
    \end{subfigure}

    \caption{Comparison of $\phi$ of the reference design and optimal solution of MSFO}\label{fig:12}
\end{figure}

To validate the optimization results, Figure~\ref{fig:12} compares the distribution of the $\phi$ at the endwall of the reference design and the optimal solution of MSFO.
Compared to the reference design, the film holes of the optimal solutions in circumferential directions are moved more close to the blade leading edge, while the films holes in axial direction are moved towards the pressure surface of the turbine blade.
Consequently, coverage area of the coolant is greatly expanded, with higher and more uniformly distributed $\phi$ values. 
With the above, the effectiveness of proposed MSFO algorithm is further demonstrated.

\section{Conclusion}

In this paper, an ensemble weighted multi-fidelity surrogate (EMFS) is proposed for more efficient turbine design optimization. Some conclusions are drawn as follows:

(1) Although the involvement of low-fidelity (LF) samples can help the multi-fidelity surrogate (MFS) to capture the global trend of the high-fidelity (HF) objective function, it can have negative effect on capturing the local trend that in small neighborhood of the true optimal solution.
Thereby, though MFS based turbine design optimization can have faster convergence rate at the beginning of the optimization process, the attained final optimal solution can be worse than the single-fidelity surrogate (SFS) based optimization that conducted with HF samples alone.

(2) Instead of purely combining the HF and LF samples to build a MFS, a SFS can be built in the region where HF samples are relatively densely distributed, which is observed to have much better prediction accuracy than that of MFS in promising areas where the true optimal solution may located. 
Then, by combining MFS and the local SFS in an ensemble weighted manner, our proposed EMFS is shown to have good accuracy in capturing the trend of the global function and that in promising local areas.

(3) Based on EMFS, a multi- and single-fidelity surrogate based optimization (MSFO) algorithm is proposed. Through tests on benchmark function, aerodynamic design optimization of turbine blade and cooling layout optimization of a turbine endwall, our proposed MSFO algorithm is observed to achieve better solutions with much faster convergence rate. Thereby, the effectiveness of our proposed MSFO algorithm has been well demonstrated.

\section*{Acknowledgments}
The authors would like to thank the anonymous referees for their valuable comments. 
This work was supported by the National Science and Technology Major Project (2019-II-0008-0028 and 2019-II-0012-0031) and the High-level Innovative and Entrepreneurial Talents Introduction Project of Qinchuangyuan (QCYRCXM-2022-210).

\bibliographystyle{asmeconf}  %
\bibliography{bibfile}%

\appendix

\end{document}